\documentclass[12pt]{article}

\usepackage{epsfig}
\usepackage{axodraw}
 
\def\ifmath#1{\relax\ifmmode #1\else $#1$\fi}%
\def\rd{\ifmath{{\mathrm{d}}}}
\newlength{\abstwidth}
\begin{document}
 
\sloppy
 
\pagestyle{empty}
 
\begin{flushright}
LU TP 01--30 \\
September 2001
\end{flushright}
 
\vspace{\fill}
 
\begin{center}
{\LARGE\bf A Simple Model for the BFKL-DGLAP}\\[3mm]
{\LARGE\bf Transition in Deep Inelastic Scattering}\\[10mm]
{\Large G\"osta Gustafson\footnote{gosta@thep.lu.se}
and  Gabriela Miu\footnote{gabriela@thep.lu.se}}\\ [2mm]
{\it Department of Theoretical Physics,}\\[1mm]
{\it Lund University, Lund, Sweden}
\end{center}

\vspace{\fill}
\begin{center}
{\bf Abstract}\\[2ex]
\begin{minipage}{\abstwidth}
A simple model is presented, which interpolates between the DGLAP and BFKL regimes in deep inelastic $ep$ scattering. The model is based on the CCFM and LDC models, and it is simple enough to provide an intuitive picture of the transition region between the two domains. Results are presented for both fixed and running coupling; for fixed coupling the transition between the domains occurs at a constant ratio between $\ln k_\perp^2$ and $\ln 1/x$, while for a running coupling it occurs for constant ratio between $\ln \ln k_\perp^2$ and $\ln 1/x$.

\end{minipage}
\end{center}

\vspace{\fill}
 
\clearpage
\pagestyle{plain}
\setcounter{page}{1}

\section{Introduction}
\label{sect-intro}

When $k_\perp$ is large and $1/x$ is limited we are in the DGLAP regime, and
when $1/x$ is large and $k_\perp$ is limited we are in the BFKL regime. An essential question is then: What is large? Where is the boundary between the regimes, and what is the behaviour in the transition region? We will here present a simple interpolating model, which can illuminate these questions. The model is based on the CCFM \cite{CCFM} and LDC \cite{LDC} models, and is simple enough to provide an intuitive picture of the dynamical mechanisms.

The model is relevant for small $x$, and has a smooth transition between large $k_\perp$, where ordered chains dominate, and small $k_\perp$, where non-ordered chains are most important. We will first study only leading terms in $\ln 1/x$, and for large $k_\perp$ the result therefore corresponds to the double log approximation and not really to the DGLAP regime. Some non-leading effects in $\ln 1/x$ will be discussed in a subsequent section. 

\begin{figure}[!b]
\begin{center}
\mbox{\epsfig{file=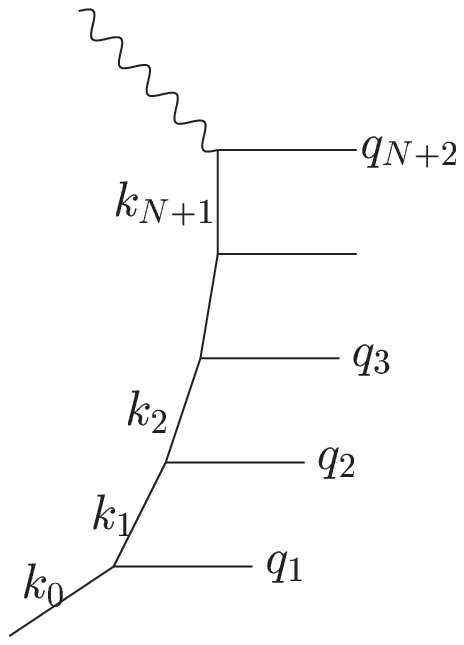,width=7.5cm}}
\end{center}
\caption{\label{fig:fan}\em A fan diagram.}
\end{figure}

For large $k_\perp$ the non-integrated structure function ${\cal F}(x,k_\perp^2)$ is dominated by contributions from ordered chains which satisfy \cite{DGLAP} (for notation see Fig~\ref{fig:fan})
\begin{eqnarray}
x & \equiv & x_{N+1} < x_N < \ldots\ x_1 < 1 \nonumber \\
k_\perp^2 & \equiv & k_{\perp N+1}^2 > k_{\perp N}^2 > \ldots\ k_{\perp 1}^2 >Q_0^2.
\end{eqnarray}
In the simpler case with a fixed coupling $\alpha_s$, and including only the $1/z$ pole in the splitting function, the contribution from each chain is a product of factors $ \frac{3\alpha_s}
{\pi} \cdot \frac{dx_i }{ x_i} \cdot \frac{d k_{\perp,i}^2 }{ k_{\perp,i}^2}$. This implies

\begin{eqnarray}
 {\cal F}(x,k_\perp^2) &\sim &\bar{\alpha} \sum_N \int \prod^N \bar{\alpha} 
\frac{dx_i}{x_i} \frac{d k_{\perp,i}^2}{k_{\perp,i}^2} 
\theta(x_{i-1} -x_{i}) \theta(k_{\perp,i}^2 -k_{\perp,i-1}^2) \nonumber \\
&= &\bar{\alpha} \sum_N \int \prod^N \bar{\alpha}\,\, dl_i \theta(l_{i} -l_{i-1})\,\, d\kappa_i \theta(\kappa_{i} - \kappa_{i-1}) \nonumber \\
{\mathrm {where}} \,\,\,\, \bar{\alpha}\!\!\!\! &\equiv &\!\!\!\! \frac{3\alpha_s}{\pi}, \,\,\,\,l \equiv \ln(1/x)\,\,\,\,{\mathrm {and}} \,\,\,\,\kappa \equiv ln(k_\perp^2/\Lambda_{QCD}^2).
\label{dglap}
\end{eqnarray}
Integration over $\kappa_i$ with the $\theta$-functions gives the phase space for $N$ ordered values $\kappa_i$. The result is $\kappa^N/N!$. (If there is a soft cutoff $k_{\perp cut}$ we obtain more exactly $(\kappa - \kappa_{cut})^N/N!$. For a fixed coupling we could use $k_{\perp cut}$ as the scale in the definition of $\kappa$, but this would be less convenient for a running coupling.) The integrations over $l_i$ give a similar result, and defining the rescaled structure function $G= 1/ \bar{\alpha} \cdot {\cal F}$, we obtain the well-known DLLA result
\begin{equation}
 G(x,k_\perp^2) \equiv \frac{1}{\bar{\alpha}} {\cal F}(x,k_\perp^2) \sim \sum_N \bar{\alpha}^N \cdot \frac{l^N}{N!} \cdot \frac{\kappa^N}{N!} = I_0(2\sqrt{\bar{\alpha} l \kappa})
\label{DGLAP}
\end{equation}
where $I_0$ is a modified Bessel function.

In the BFKL region non-ordered chains contribute, and the result is a power-like increase for small $x$-values \cite{BFKL},

\begin{equation}
 {\cal F} \sim \frac{1}{x^\lambda} f(\kappa,l)
\end{equation}
where the function $f(\kappa,l)$ describes a random walk in $\kappa$ \cite{BFKL,cigar}. 

In the interpolation region we must calculate suppressed
contributions from non-ordered chains. It is then necessary to specify
the separation between initial state radiation and final state
radiation, which is not given by Nature, but has to be defined by the
calculation scheme. This separation should have the property that final 
state emissions do not affect the total cross section, i. e. the structure 
function, and their effect on the final state can be described by Sudakov 
form factors. A particular scheme was chosen by Ciafaloni,
Catani, Fiorani, and Marchesini (the CCFM model) \cite{CCFM}. Here the
initial state radiation is ordered in {\em angle} (or rapidity) and
{\em energy} (or $q_+=q_0 + q_L$). The contribution from a particular chain is
then given by the expression
\begin{equation}
 \prod \bar{\alpha}\frac{dz_i}{z_i} \frac{d q_{\perp,i}^2}{q_{\perp,i}^2} \Delta_{ne}(z_i,k_{\perp,i},q_{\perp,i})
\label{CCFM}
\end{equation}
where $\Delta_{ne}$ is a noneikonal form factor and $z_i=x_i/x_{i-1}$.

The Linked Dipole Chain (LDC) model \cite{LDC} is a
reformulation and generalization of the CCFM result in a scheme, where
more gluons are treated as final state radiation. The initial chain is
ordered in $q_+=q_0 + q_L$ {\em and} $q_-=q_0 - q_L$, and $q_\perp$
satisfies $q_\perp > \min(k_{\perp,i},k_{\perp,i-1})$. This implies
that there are fewer chains; one LDC chain corresponds to a set of
CCFM chains. It then turns out that all the corresponding noneikonal
form factors add up to just unity, and thus the contribution from each
such chain is given by the simple product
\begin{equation}
\prod \bar{\alpha} \frac{\rd z_i}{z_i} 
\frac{\rd q_{\perp,i}^2}{q_{\perp,i}^2} .
\label{vikt}
\end{equation}
We note in particular that this expression is totally left-right
symmetric, meaning that we get the same result if we start the chain
in the photon end, instead of in the proton end. (This implies that the formalism automatically takes into account contributions from ``resolved photons''.)

We can express this result in the link momenta $k_i$, instead of the
final state momenta $q_i$. For ordered chains we have $q_{\perp,i} \approx k_{\perp,i}$, and thus for these chains the expression in eq.~(\ref{vikt}) agrees with eq.~(\ref{dglap}), but for non-ordered chains this is no longer the case. Using the relations $\rd^2 q_{\perp,i} = \rd^2
k_{\perp,i}$ and $q_{\perp,i}^2 \approx \max( k_{\perp,i}^2,
k_{\perp,i-1}^2)$ we find the following weights
\begin{eqnarray} 
\frac{\rd^2 q_{\perp,i}}{q_{\perp,i}^2} \approx \frac{\rd^2 k_{\perp,i}}
{k_{\perp,i}^2}
\,\,\,\,{\mathrm {for}}\,\,\,\, k_{\perp,i} > k_{\perp,i-1} \, \, {\mathrm {and}}\nonumber \\
\frac{\rd^2 q_{\perp,i}}{q_{\perp,i}^2} \approx \frac{\rd^2 k_{\perp,i}}
{k_{\perp,i}^2} 
\cdot \frac{ k_{\perp,i}^2}{k_{\perp,i-1}^2}\,\,\,\,{\mathrm {for}}\,\,\,\,
k_{\perp,i} < k_{\perp,i-1}.
\label{step-up-down}
\end{eqnarray}
Thus, for a step down in $k_\perp$ we have an extra suppression factor 
$ k_{\perp,i}^2/k_{\perp,i-1}^2$, which reduces the weight for non-ordered chains. (This suppression also implies that if the chain 
goes up to $k_{\perp,max}$ and then down to $k_{\perp,final}$ we obtain a 
factor $1/k_{\perp,max}^4$, which corresponds to the cross section for a 
hard parton-parton subcollision.)

In the following we will first discuss a model for a fixed coupling, and after that the more relevant situation with a running $\alpha_s$. In section 4 we discuss some effects of non-leading terms in $\ln 1/x$. We end by some comments on the Laplace transforms in section 5.

\section{Fixed coupling}

As discussed in the introduction, in the LDC and CCFM models downward steps in $k_\perp$ are suppressed by the factor $ k_{\perp,i}^2 / k_{\perp,i-1}^2$ in eq.~(\ref{step-up-down}). Expressed in the logarithmic variable $\kappa$, large downward steps are thus suppressed by an exponential factor $\exp(\kappa_i - \kappa_{i-1})$. This implies that
the effective allowed distance, $\delta$, for a downward step is given by the average value of this exponential, which means
\begin{equation}
\delta \sim \int_{\kappa_i} (\kappa_{i-1} - \kappa_i) e^{-(\kappa_{i-1} - \kappa_i)} \rd\kappa_{i-1} = 1.
\label{delta}
\end{equation}
This increases the phase space; the boundary is effectively given by $\kappa_{i} > \kappa_{i-1} - \delta$ instead of the strict ordering $\kappa_{i} > \kappa_{i-1}$ in eq.~(\ref{dglap}) relevant in the DGLAP limit. (This has similarities with the van der Waals gas formula, where the gas molecules take up a fixed volume. Here the corresponding ``volume'' is negative, which thus increases the phase space.) The result is that the phase space factor $\frac{\kappa^N}{N!}$ is replaced by $\frac{(\kappa+N\delta)^N}{N!}$ \footnote{This result is exactly correct only if the non-ordered variables $\kappa_i$ are allowed to take on negative values in the middle of the chain. If all $\kappa_i$ are restricted to be positive, we get instead $(\kappa+\delta)[\kappa+(N+1)\delta]^{N-1}/N!$. This small difference does not modify our conclusions.}. Thus we obtain

\begin{equation}
G(x,k_\perp^2) \sim \sum_N \frac{(\bar{\alpha} l)^N}{N!} \frac{(\kappa+N\delta)^N}{N!}.
\label{fixsumma}
\end{equation}

When {\em $\kappa$ is very large} we can neglect the term $N \delta$, and thus obtain the DGLAP result from eq.~(\ref{DGLAP}): 
\begin{equation}
G(x,k_\perp^2) \sim \sum_N \frac{(\bar{\alpha} l \kappa)^N}{(N!)^2} = I_0(2\sqrt{\bar{\alpha} l \kappa}) \approx \frac{\exp(2 \sqrt{\bar{\alpha} l \kappa})}{\sqrt{4 \pi \sqrt{\bar{\alpha} l \kappa}}}. 
\label{dglapasfix}
\end{equation}
The last approximation is relevant when the argument of $I_0$ is large compared to 1.

When {\em $\kappa$ is small}, it can instead be neglected compared to $N \delta$, and we obtain using Sterling's formula
\begin{eqnarray}
G\!\! & \sim & \!\!\sum_N \frac{(\bar{\alpha} l)^N}{N!} \frac{(N\delta)^N}{N!} \approx \sum_N \frac{(\bar{\alpha} l)^N}{N!} \delta^N \frac{e^N}{\sqrt{2 \pi N}} \nonumber \\
& \approx & \frac{ \exp(\bar{\alpha} \delta e l)}{\sqrt{2 \pi \bar{\alpha} \delta e l}} = \frac{1}{\sqrt{2 \pi \lambda \ln1/x}} \cdot \frac{1}{x^\lambda} \nonumber \\
{\mathrm {with}} \,\,\,\,\lambda \!\!& = & \!\!e \delta \bar{\alpha}.
\label{bfklasfix}
\end{eqnarray}
Thus for $\delta = 1$ we get $\lambda = e \bar{\alpha} = 2.72\bar{\alpha}$. This should be compared with the leading order result for the BFKL equation, $\lambda = 4 \ln 2 \, \bar{\alpha} = 2.77\bar{\alpha}$.

We now also want to study the behaviour of the sum in eq~(\ref{fixsumma}) in the {\em transition region} between the two extreme situations. This can be estimated by the saddle point method. Using Sterling's formula we get from eq.~(\ref{fixsumma})

\begin{equation}
G \sim \sum_N \exp\left\{N[\ln(\bar{\alpha} l) + \ln(\kappa + N\delta) - 2\ln N + 2]\right\} \equiv \sum_N \exp\{h(N)\}.
\end{equation}
In this sum the terms with $N$-values around $N_{max}$ dominate, where $N_{max}$ is the solution to the equation 
\begin{equation}
h'(N_{max}) = \ln(\bar{\alpha} l) + \ln(\kappa + N_{max} \delta) - 2\ln N_{max} + \frac{N_{max} \delta }{\kappa + N_{max} \delta} =0.
\label{Nmax}
\end{equation}

Let us first assume that $k_\perp$ is large and satisfies the inequality $N_{max} \delta < \kappa$. This implies that for the dominating terms in the sum we have $\kappa + N \delta \approx \kappa$, which corresponds to the ``DGLAP region'' as discussed above. Inserted into eq.~(\ref{Nmax}) this implies $N_{max} \approx \sqrt{\bar{\alpha} l \kappa} + \bar{\alpha} l \delta$. For consistency with our assumption, we must then demand this expression to be smaller than $\kappa / \delta$, which implies $\kappa  > \bar{\alpha} \frac{3+\sqrt{5}}{2} \delta^2 l$. For $\delta = 1$ we thus obtain

\begin{equation}
\kappa > \frac{3+\sqrt{5}}{2}\,\bar{\alpha}\, l \approx 2.62 \,\bar{\alpha}\, l.
\label{DGLAPlimit}
\end{equation} 

Next we assume that $N_{max} \delta > \kappa$. This implies that for the dominating terms in the sum we have $\kappa + N \delta \approx N \delta$, which corresponds to the BFKL region. Eq.~(\ref{Nmax}) now gives $N_{max} \approx \bar{\alpha}\, l \,\delta \,e$. Consistency with the assumption, $N_{max} > \kappa / \delta$, now demands $\kappa < \bar{\alpha}\, e\, \delta^2 l$. For $\delta = 1$ we therefore obtain in this case

\begin{equation}
\kappa < \bar{\alpha}\, e\, l \approx 2.72\, \bar{\alpha}\, l.
\label{BFKLlimit}
\end{equation}
This is very close to the limit in eq.~(\ref{DGLAPlimit}). Consequently this line in the $(l,\kappa)$ plane corresponds to the transition between the BFKL and DGLAP regions. Due to the result for the exponent $\lambda$ in eq.~(\ref{bfklasfix}) this line can be written $\kappa/l\approx \lambda$.

For small $x$ we can also go one step further, and expand eq.~(\ref{Nmax}) in powers of $\kappa/l$. This gives $N_{max}\approx e \delta \bar{\alpha} l + \kappa^2 /(2  e \delta^3 \bar{\alpha} l) - 1$, which implies (with $\lambda$ still given by $e \delta \bar{\alpha}$)
\begin{equation}
G \sim \frac{1}{\sqrt{2 \pi \lambda l}} \exp \left[\lambda l + \frac{\kappa}{\delta} - \frac{\kappa^2}{2 \delta^2 \lambda l}\right].
\label{Gauss}
\end{equation}
We recognize here the Gaussian distribution with a width proportional to $\lambda l$, corresponding to a random walk in $\kappa$. This can be compared with the result of the BFKL equation \cite{cigar}:
\begin{equation}
G \sim \frac{1}{\sqrt{2 \pi \lambda l}} \exp \left[\lambda l + \frac{\kappa}{2} - \frac{\kappa^2}{\Delta \lambda l} \right]; \,\,\,\, \Delta = \frac{14 \zeta(3)}{\ln2} \approx 24.3.
\end{equation}
We see that for $\delta \approx 1$ the distribution in eq.~(\ref{Gauss}) is significantly more narrow. This is perhaps not surprising as the exponential suppression for large values of $\kappa_{i-1} - \kappa_i$ in eq.~(\ref{delta}) is replaced by the sharp cut for $\kappa_{i} = \kappa_{i-1} - \delta$. The model is, however, only intended to illustrate the qualitative features and not to give a quantitative description.

\begin{figure}[t]
\begin{center}
\mbox{\epsfig{file=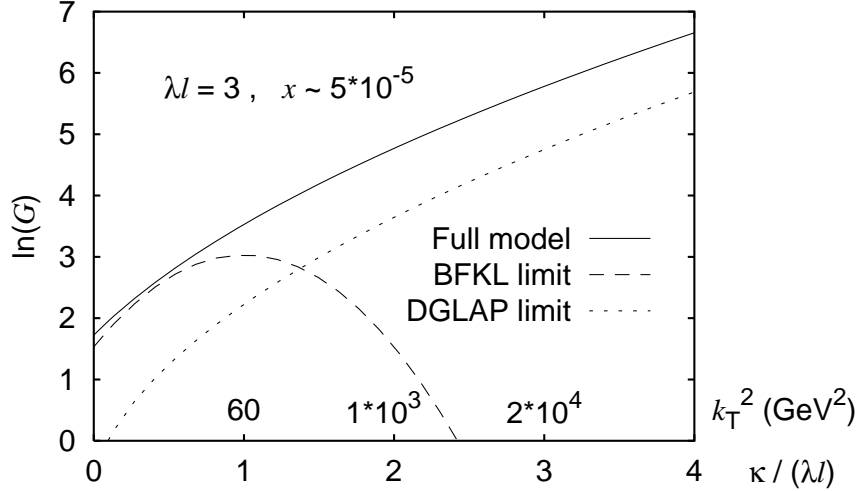,width=10cm}}
\end{center}
\caption{\label{fig:fix}\em The logarithm of the rescaled structure function obtained for a fixed coupling, for $\lambda l = 3$, which corresponds to $x\sim 5\cdot 10^{-5}$.
The approximations for small and large $k_\perp$-values are indicated by the dashed and doted lines respectively. The horizontal axis is the combination $\kappa/\lambda l$, chosen so that the transition occurs when this variable has values around 1. Corresponding values of $k_{\perp}^2$ in $GeV^2$ are also indicated.}
\end{figure}

The result of a numerical evaluation of eq.~(\ref{fixsumma}) is presented in Fig.~\ref{fig:fix} together with the DGLAP and BFKL approximations of eqs.~(\ref{dglapasfix}) and (\ref{Gauss}). 
We have here chosen $\delta=1$. The result depends only on the two combinations $\lambda l$ and $\kappa/\lambda l$, where the last combination is chosen so that the transition occurs when this variable is around 1. 
We see in this figure how our result interpolates between the BFKL and DGLAP expressions, although the normalization of the DGLAP approximation is off by roughly a factor $e$, even for large values of $k_\perp$. An essential feature is also that $G$ grows monotonically with $\kappa$. The Gaussian behaviour is only obtained in an expansion of $\ln {\cal F}$ to second order in powers of $\kappa/l$.
%

\subsection*{Conclusion:}

The expression ${\cal F}(x,k_\perp^2) \sim \bar{\alpha} \sum_N \frac{(\bar{\alpha} l)^N}{N!} \frac{(\kappa+N\delta)^N}{N!}$ with $\delta \approx 1$ reproduces DGLAP evolution (or more correctly the double log approximation $\sim \exp (2\sqrt{\bar{\alpha} l \kappa})$) for large values of $\kappa/l$, and reproduces BFKL evolution $\sim x^{-\lambda}$ with $\lambda \approx e \bar{\alpha}$ for smaller $\kappa/l$.
The  transition occurs for a fixed ratio between $\kappa$ and $l$:
\begin{equation}
\frac{\kappa}{l} =  \frac{\ln k_\perp^2}{\ln 1/x} \approx \lambda \approx e \,\bar{\alpha}.
\label{fixgrans}
\end{equation}
We also note that the structure function grows monotonically with increasing $\ln k_\perp^2$, in contrast to the Gaussian distribution in the BFKL approximation, related to a random walk in $\ln k_\perp^2$. The Gaussian distribution is only obtained in a second order expansion in $\kappa/l$.

\section{Running Coupling}

For a running coupling we use the notation
\begin{eqnarray}
\bar{\alpha} & = & \frac{3\alpha_s}{\pi} \equiv \frac{\alpha_0}{\kappa} \nonumber \\
u & \equiv & \ln (\kappa/\kappa_{cut}) = \ln \ln (k_{\perp}^2/\Lambda^2) - \ln \ln (k_{\perp cut}^2/\Lambda^2)\nonumber \\
G & \equiv & \frac{\kappa}{\alpha_0} \cdot {\cal F}.
\label{defGu}
\end{eqnarray}
As we only include purely gluonic chains, it would be most consistent to use the value for $\alpha_s$ obtained for no flavours, which corresponds to $\alpha_0=12/11\approx1$. In eq.~(\ref{vikt}) the scale in $\alpha_s$ at a certain vertex should be given by the largest of the associated virtualities, which means $\max (k_{\perp,i}^2, k_{\perp,i-1}^2)$. For a step up this is $k_{\perp,i}^2$, and thus the running coupling contains a factor $1/\kappa_i$. For these steps we also have $q_\perp \approx k_\perp$, and thus the weight in eq.~(\ref{vikt}) becomes
\begin{equation}
\prod \frac{\alpha_0}{\kappa_i} \rd l_i \rd \kappa_i = \prod \alpha_0 \rd l_i \rd u_i .
\label{viktrunningup}
\end{equation}
In the DGLAP region $u$ is large, and downward steps can be neglected. In analogy with eqs.~(\ref{DGLAP}) and (\ref{dglapasfix}) we then find the well-known result
\begin{equation}
G  \sim \sum \alpha_0^N \frac{l^N}{N!} \cdot \frac{u^N}{N!} = I_0(2\sqrt{\alpha_0 l u}),
\label{gdglap} 
\end{equation}
which resembles eq.~(\ref{dglapasfix}) for fixed coupling, only with $\ln k_\perp^2$ replaced by $\ln\ln k_\perp^2$.

Including downward steps we see, however, that the effect of the suppression factor $ k_{\perp,i}^2 / k_{\perp,i-1}^2$ becomes quite different for a running coupling compared to the case with a fixed coupling. Expressed in the variables $u_i$, this factor is
\begin{equation}
\exp[-(e^{u_{i-1}} -e^{u_i})] \sim \exp[-e^{u_{i-1}}(u_{i-1}-u_i)].
\end{equation}
When $u_i$ and $u_{i-1}$ are large, this factor drops rapidly with increasing difference $u_{i-1} - u_i$.
This implies that it is easy to go down when $u_i$ and $u_{i-1}$ are small, but very difficult when they are large. A consequence of this feature is the result observed in ref.  \cite{jimhamid}, that a typical chain contains two parts. In the first part the $k_\perp$-values are relatively small, and it is therefore easy to go up and down in $u$ (or $k_{\perp}$), and non-ordered $k_\perp$-values are important. The second part is an ordered, DGLAP-type, chain, where $k_\perp$ increases towards its final value. In ref. \cite{hamid} we also found that for large $l$ (small $x$) an approximate solution to the LDC equation has the factorized form
\begin{equation}
G  \sim \frac{1}{x^\lambda} \cdot \kappa^{\frac{\alpha_0}{\lambda}}.
\label{Gfactor} 
\end{equation}

Let us study a chain with $N$ links, out of which $N-k$ correspond to the first part with small non-ordered $k_\perp$, and the remaining $k$ links belong to the second part with increasing $k_\perp$. Assume that the effective space for each $u_i$ in the soft part is given by a quantity $\Delta$, and we will below estimate its magnitude. The total weight for this part then becomes $\Delta^{N-k}$. For the $k$ links in the second, ordered, part the phase space becomes as before $u^k/k!$. Thus the total result is 

\begin{equation}
G  \sim \sum_N \frac{(\alpha_0 l)^N}{N!} \sum_{k=0}^N \frac{u^k}{k!} \Delta^{N-k} 
=\sum_N \frac{(\alpha_0 l \Delta)^N}{N!} \sum_{k=0}^N \frac{(u/\Delta)^k}{k!}. 
\label{Gslut} 
\end{equation}

We first study the behaviour of this expression in the two extreme situations with small and large $k_\perp$-values.
\vspace{5mm}

{\em i) BFKL limit, $u$ small.}

Assume first that $u/\Delta$ is small compared to dominating values of $N$. Then the sum over $k$ gives approximately $\exp(u/\Delta)$. Thus $G$ takes the form

\begin{equation}
G  \sim  \sum_N \frac{(\alpha_0 l \Delta)^N}{N!} \exp(u/\Delta) = \exp(\alpha_0 l \Delta + u/\Delta) = \frac{1}{x^\lambda} \cdot \kappa^{\alpha_0 / \lambda}
\label{Gbfkl} 
\end{equation}

\begin{equation}
\mathrm{where} \,\,\,\, \lambda = \alpha_0 \Delta.
\label{lambda}
\end{equation}
We note that this result has just the factorized form of eq.~(\ref{Gfactor}), which was obtained in ref. \cite{hamid}.
From eq.~(\ref{Gbfkl}) we see that our assumptions imply that the sum over $N$ is dominated by $N$-values around $\alpha_0 l \Delta $. For consistency we therefore demand $u/\Delta < \alpha_0 l \Delta$ or 
\begin{equation}
u < \alpha_0 l \Delta^2 = \frac{\lambda^2 l}{\alpha_0}. 
\label{bfklgrans}
\end{equation}
\vspace{5mm}

{\em ii)  DGLAP limit, $u$ large.}

We now assume that $u$ is so large that $u/\Delta$ is larger than $N$ for the most important values of $N$. The sum over $k$ is then dominated by its last term, $\frac{(u/\Delta)^N}{N!}$,  which gives the result in eq.~(\ref{gdglap}):
\begin{equation}
G  \sim  \sum_N \frac{(\alpha_0 l u)^N}{(N!)^2} = I_0(2\sqrt{\alpha_0 l u}).
\label{Gdglap} 
\end{equation}

Using Sterling's formula we find that in this case the sum is dominated by $N$-values around $N_{max} = \sqrt{\alpha_0 l u}$. For consistency we therefore now demand 
\begin{equation}
u/\Delta > \sqrt{\alpha_0 l u} \,\,\, \mathrm{or} \,\,\, u > \alpha_0 l \Delta^2 = \frac{\lambda^2 l}{\alpha_0}.
\label{dglapgrans}
\end{equation}

 We see that this limit
 coincides exactly with the BFKL limit 
in eq.~(\ref{bfklgrans}). Thus the regions of validity for the BFKL and DGLAP approximations of eqs.~(\ref{Gbfkl}) and (\ref{Gdglap}) respectively join at the common boundary for the fixed ratio between $u$ and $l$:

\begin{equation}
u / l  \approx \lambda ^2 / \alpha_0.
\label{grans}
\end{equation}

\begin{figure}[!t]
\begin{center}
{
\mbox{\epsfig{file=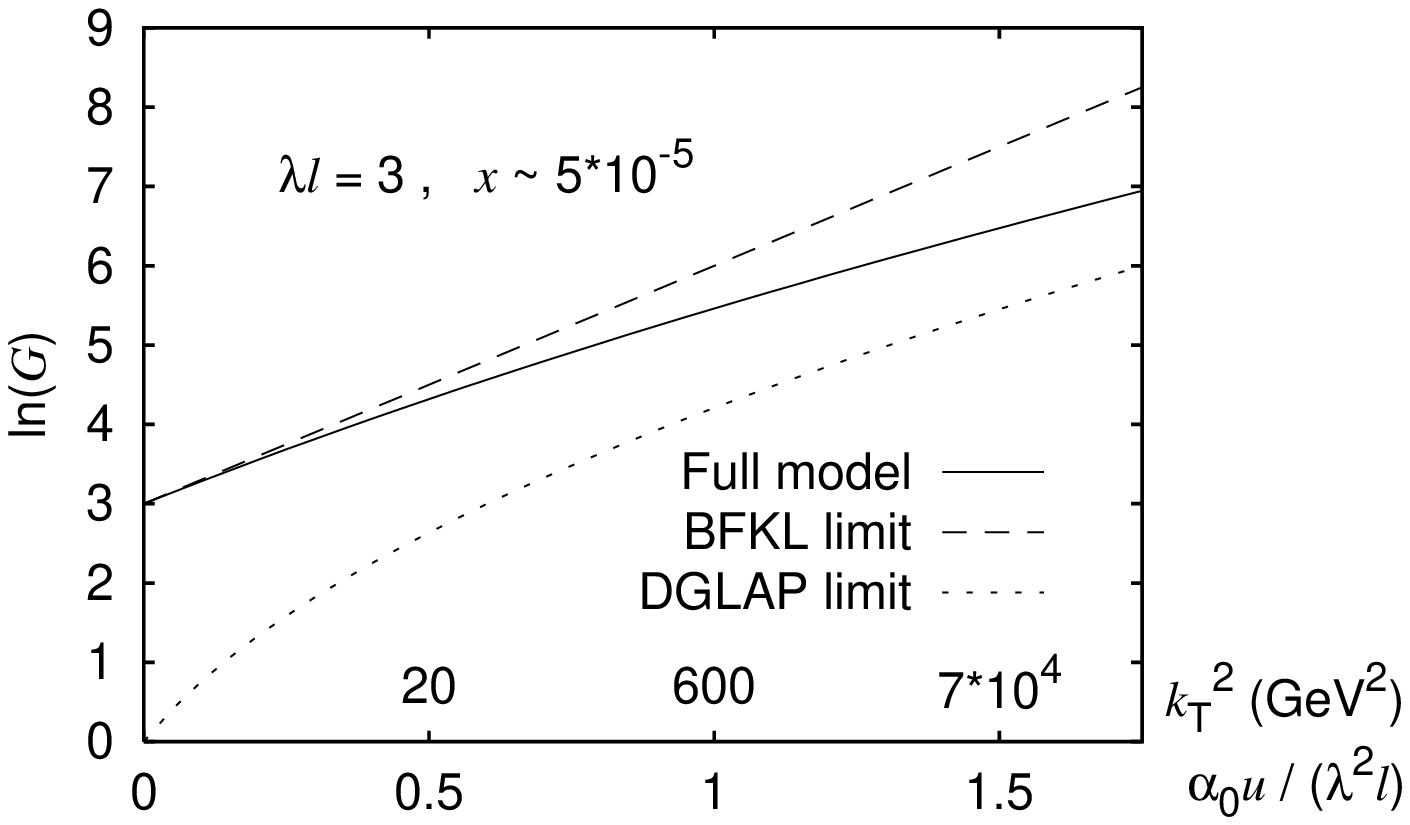, width=10cm}}
{\mbox{\begin{picture}(400,10)(0,0)\Text(200,5)[]{(a)} \end{picture}}}
\mbox{\epsfig{file=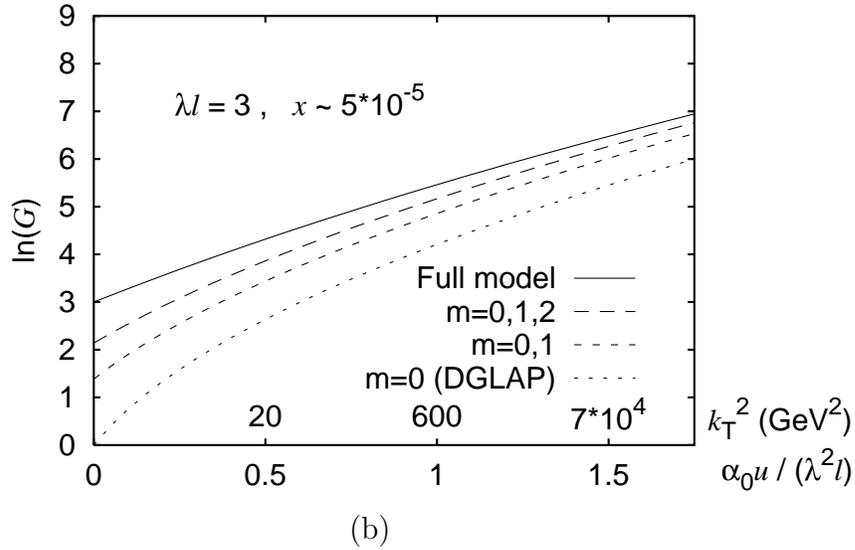, width=10cm}}
{\mbox{\begin{picture}(400,10)(0,0)\Text(200,5)[]{(b)} \end{picture}}}
}
\end{center}
\caption{\label{fig:run}\em (a) The logarithm of the rescaled structure function for a running coupling. $\lambda l = 3$, which corresponds to $x\sim 5\cdot 10^{-5}$. 
The approximations for small and large $k_\perp$-values are indicated by the dashed and doted lines respectively. The horizontal axis is the combination $\alpha_{0} u/( \lambda^2 l)$, chosen so that the transition occurs when this variable has values around 1. Corresponding values of $k_{\perp}^2$ in $GeV^2$ are also indicated. (b) The contributions from the first three terms in eq.~(\ref{Imsumma}). Note that $m=0$ corresponds to the DGLAP limit. }
\end{figure}

These results are confirmed by a numerical evaluation of eq.~(\ref{Gslut}) presented in Fig.~\ref{fig:run}.
Introducing the variables $v = \lambda l$ and $w= \frac{\alpha_0 u}{\lambda^2 l}$ we have 
\begin{eqnarray}
& \ln G  \sim  v(1+w)\,;\,\,\,\, & w  < 1 \,\,(BFKL) \nonumber \\
& \ln G  \sim   \ln I_0(2 v \sqrt {w}) \approx 2 v \sqrt {w} - { \frac{1}{2}} \ln (4 \pi v \sqrt{w})\,;\,\,\,\, & w > 1 \,\,(DGLAP).
\label{lnapprox}
\end{eqnarray}
The variables $v$ and $w$ are chosen so that the transition occurs around $w=1$. Note that the asymptotic approximation for the Bessel function,
$ I_0(z) \approx \exp(z) / \sqrt{2 \pi z} $,
 is relevant in all of the
 interesting kinematic region. The model should be applied to small $x$. For $x<0.03$ and $\lambda \sim 0.3$ we have $\lambda l > 1$, which together with the constraint in eq.~(\ref{dglapgrans}) implies that $\alpha_0 l u > 1$. Thus the argument in the Bessel function is here always larger than 2, and for these values the error in the approximation is less than 10\%. 

It is also interesting to rewrite eq.~(\ref{Gslut}) in the following form, where $m=N-k$:
\begin{equation}
G  \sim \sum_m (\alpha_0 l \Delta)^m \sum_{N=m}^\infty \frac{(\alpha_0 l u)^{N-m}}{N! (N-m)!} 
=\sum_m \left( \frac{\alpha_0 l \Delta^2}{u} \right)^{\frac{m}{2}} I_m(2\sqrt{\alpha_0 l u}). 
\label{Imsumma} 
\end{equation}
Here $m$ is the number of steps in the initial non-ordered part, and $I_m$ are modified Bessel functions. The DGLAP approximation in eq. (\ref{Gdglap}) corresponds to just the first term with $m=0$, and Fig.~\ref{fig:run}b shows how this series approaches the full result when more terms are included.
\vspace{5mm}

{\em Estimate of the effective phase space $\Delta$.} 

\begin{figure}[t]
\begin{center}
\mbox{\epsfig{file=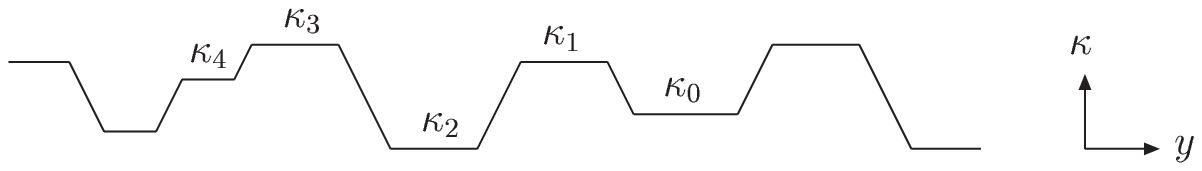}}
\end{center}
\caption{\label{fig:steps}\em  The initial part of a  chain dominated by alternating steps up and down in $\kappa$.}
\end{figure}

We now want to study in more detail the initial part of the chain, where $k_\perp$ is small, $\alpha_s$ large and where it therefore is easy to go up and down in $k_\perp$. If  $\alpha_s$ is assumed to be proportional to $1/\kappa$, the result is sensitive to the infrared region, and thus depends on the effective cutoff for small $k_\perp$ \cite{hamid, cigar}. Experience from the LDC MC shows that long chains often have the form illustrated in Fig.~\ref{fig:steps}, with alternating steps up and down in $k_\perp$. Occasionally there are also two steps in the same direction (e. g. the step $\kappa_4$ in Fig.~\ref{fig:steps}). Neglecting such ``stairs'' will underestimate the power $\lambda$ but, as we will see, qualitatively it does reproduce the dependence of $\lambda$ on the cutoff.

Let us study two typical steps, a step up to a high level $\kappa_1$, followed by a step down to a low level $\kappa_2$ (cf. Fig.~\ref{fig:steps}). (For simplicity we use the indices 1 and 2, even if these steps are not in the beginning of the chain.) The definition of $\Delta$ implies that the $\kappa$-integrations in these two steps (including the factors $\bar{\alpha}(\kappa)$) should correspond to a factor $(\alpha_0 \Delta)^2$. In each vertex the argument in $\alpha_s$ is given by the largest of the adjacent $k_\perp$-values. This implies in this case two factors $\alpha_0 / \kappa_1$ and no factor $\alpha_0 / \kappa_2$. (Due to the singular behaviour of $\alpha_s$, we should not neglect the difference between $\alpha_0/\kappa_1$ and $\alpha_0/\kappa_2$ for these low $\kappa$-values.) Neglecting constraints other than the fact that $\kappa_1$ must be larger than $\kappa_2$, we therefore get as a first estimate the relation
\begin{equation}
(\alpha_0 \Delta)^2 = \int_{\kappa_c}^\infty \left(\frac{\alpha_0}{\kappa_1}\right)^2 d\kappa_1 d\kappa_2 \theta(\kappa_1 - \kappa_2) \exp[-(1+\lambda)(\kappa_1 - \kappa_2)].
\label{Deltaapprox}
\end{equation}
Here $\kappa_c \equiv \ln(k_{\perp cut}^2/\Lambda^2)$. A factor $\exp[-(\kappa_1 - \kappa_2)]$ corresponds to the factor $k_{\perp,i}^2 / k_{\perp,i-1}^2$ in eq. (\ref{step-up-down}) when $k_\perp$ is decreasing. 
A further factor $\exp[-\lambda(\kappa_1 - \kappa_2)]$ follows because $\kappa_1 - \kappa_2$ is the minimum step in $l=\ln 1/x$, and in this interval there can be no other vertices. Without such a constraint $G$ would in this $l$-interval grow by a factor $\exp[\lambda(\kappa_1 - \kappa_2)]$, and the constraint therefore corresponds to the compensating factor $\exp[-\lambda(\kappa_1 - \kappa_2)]$ in the integrand. 

In the estimate in eq.~(\ref{Deltaapprox}) the constraints $ \kappa_1 > \kappa_0$ and $ \kappa_2 < \kappa_3$ are not taken into account. Most important is here the first inequality. Due to the factor $1/\kappa_1^2$, small $\kappa_1$-values give a large contribution to the integral in eq. (\ref{Deltaapprox}), which is reduced when the constraint $ \kappa_1 > \kappa_0$ is taken into account.
Integrating over $\kappa_1$, we obtain from eq. (\ref{Deltaapprox}) also the normalized distribution, $P(\kappa_2)$, in $\kappa_2$:
\begin{equation}
P(\kappa_2) = \frac{1}{e^{\widehat{\lambda} \kappa_c} E_1(\widehat{\lambda} \kappa_c)}\left\{\frac{1}{\kappa_2} - \widehat{\lambda} e^{\widehat{\lambda} \kappa_2} E_1(\widehat{\lambda} \kappa_2) \right\}
\label{Pkappa2}
\end{equation}
where the function $E_1$ is an exponential integral and $\widehat{\lambda} \equiv 1+\lambda$.
Within our approximation, this expression also describes the distribution in the earlier ``low step'' $\kappa_0$. Thus the constraint $ \kappa_1 > \kappa_0$ gives an extra weight (where $P$ is given by eq. (\ref{Pkappa2}))
\begin{equation}
Prob(\kappa_0 < \kappa_1) = \int_{\kappa_c}^{\kappa_1} P(\kappa_0) d \kappa_0 = 1 - \frac{e^{\widehat{\lambda} \kappa_1} E_1(\widehat{\lambda} \kappa_1)}{e^{\widehat{\lambda} \kappa_c} E_1(\widehat{\lambda} \kappa_c)}.
\label{Prob(kappa_0 < kappa_1)}
\end{equation}
Including this factor in the integrand in eq. (\ref{Deltaapprox}) we obtain (after integrating over $\kappa_2$) the relation
\begin{equation}
\Delta^2 = \int_{\kappa_c}^{\infty} \left(\frac{1}{\kappa_1}\right)^2 d\kappa_1 \left[1 - \frac{e^{\widehat{\lambda} \kappa_1} E_1(\widehat{\lambda} \kappa_1)}{e^{\widehat{\lambda} \kappa_c} E_1(\widehat{\lambda} \kappa_c)}\right] \frac{1}{\widehat{\lambda}} \left[1 - e^{-\widehat{\lambda}(\kappa_1 - \kappa_c)} \right].
\label{Deltaint}
\end{equation}
It would be possible to include a similar factor representing the constraint $\kappa_2 < \kappa_3$, and then, in an iterative scheme, also use the improved distributions to find new weights. This procedure would, however, be algebraically quite complicated, and not motivated in view of the qualitative nature of our approximation.

\begin{figure}[t]
\begin{center}
\mbox{\epsfig{file=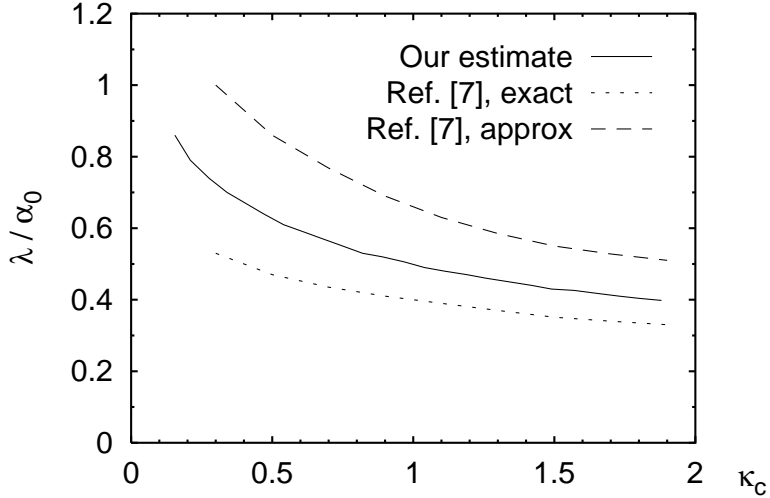,width=10cm }}
\end{center}
\caption{\label{fig:delta}\em The effective phase space $\Delta=\lambda/\alpha_{0}$ as a function of $\kappa_{cut}= \ln(k_{\perp cut}^2/\Lambda^2)$. The estimate from eq.~(\ref{Deltaint}) is shown together with the results from solving the LDC integral equation in reference \cite{hamid}. Here the upper (dashed) line is obtained using the approximations in eq.~(\ref{step-up-down}), which are also used in our model.}
\end{figure}

A numerical evaluation of the integral in eq. (\ref{Deltaint}) gives the relation between $\Delta = \lambda/\alpha_0$ and $\kappa_c$ presented in Fig.~\ref{fig:delta}.\footnote{We note that the integral in eq. (\ref{Deltaint}) is a function only of the product $\widehat{\lambda} \kappa_c$. Thus the relation between $\Delta$ and $\kappa_c$ is easily obtained by calculating the integral for different values of $\widehat{\lambda} \kappa_c$. This gives directly $\Delta$ and its corresponding value for $\kappa_c = \widehat{\lambda} \kappa_c/(1+\Delta \alpha_0)$.} In this figure we also show the exact result presented in ref. \cite{hamid}, obtained by solving an integral equation which includes the effect from all possible chains. The dotted line is obtained using a more accurate integration over azimuth angles, while the dashed line is obtained using the approximations in eq.~(\ref{step-up-down}), and therefore corresponds more directly to our present model. We see that the result from eq.~(\ref{Deltaint}) indeed well reproduces the qualitative behaviour of the full solution. The neglect of ``double steps'' in the chains implies a somewhat lower value for $\Delta$ (our result lies between the two curves from ref. \cite{hamid}), but the qualitative agreement gives further support to our assumption that this type of chains gives the dominant contribution for small $x$, for both large and small $k_\perp^2$.
%

\subsection*{Conclusion:}

For a running coupling it is easy to go up and down in $k_\perp$, only as long as $k_\perp$ is small and $\alpha_s$ large. Therefore dominating chains contain a non-ordered part with small $k_\perp$-values, and an ordered part, where $k_\perp$ increases towards its final value. If the effective phase space for the variables $u_i = \ln k_{\perp i}^2 $ in the non-ordered part is denoted $\Delta$, then the expression in eq.~(\ref{Gslut}) interpolates smoothly between the DGLAP and BFKL regions. The BFKL exponent is given by $\lambda=\alpha_0 \Delta$, and the transition between the regimes occurs for a fixed ratio between  $u=\ln \ln k_\perp^2$ and $l=\ln(1/x)$,
$u / l  \approx \lambda ^2 / \alpha_0$.

\section{Non-leading contributions}

The replacement $\kappa^N \rightarrow (\kappa+N\delta)^N$ in eq.~(\ref{fixsumma}) includes the effect of non-leading terms in $\ln k_\perp^2$. In ref. \cite{LDC} it is discussed how a similar modification of the factor $l^N$ can account for some non-leading $\ln 1/x$ effects.

It is well-known that non-leading contributions have a significant influence on the value of the BFKL exponent $\lambda$ \cite{LDC, non-leading}. The $1/z$ pole in the splitting function gives the leading contribution to the growth of the structure functions for small $x$. The contribution from the $1/(1-z)$ pole is compensated by a Sudakov form factor \cite{CCFM}. This is related to the fact that, when a gluon is split into two gluons with energy fractions $z$ and $1-z$ with $z<<1$, then the softer one can be regarded as a new emission, while the harder one (with energy $1-z$) corresponds to the original parent gluon, only with slightly reduced energy. Instead of including the $1/z$ pole for $0<z<1$ and neglecting the  $1/(1-z)$ pole, it would also be possible to take both poles into account, but restrict the $z$-interval to $0<z<0.5$. This gives approximately the same result due to the relation

\begin{equation}
\int_\epsilon^1 \frac{1}{z} dz = \int_\epsilon^{0.5} \left[\frac{1}{z} +\frac{1}{1-z} \right] dz = \ln 1/\epsilon.
\label{epsilon}
\end{equation}
Including the non-singular terms in the splitting function
\begin{equation}
P_{gg} \propto \frac{1}{z} +\frac{1}{1-z} - 2 + z(1-z) 
\label{Pgg}
\end{equation}
reduces this integral. Integrating from $\epsilon$ to 0.5 gives
\begin{equation}
\int_\epsilon^{0.5} P_{gg}(z) dz \propto \ln 1/\epsilon - \frac{11}{12}.
\label{lgrans}
\end{equation}
In analogy with eq.~(\ref{epsilon}) we get exactly the same result if we split $P_{gg}$ in the following way in two symmetric parts, singular at $z=0$ and $z=1$ respectively
\begin{equation}
P_{gg} \propto \frac{1 + (1-z)^3}{2z} +\frac{1 + z^3}{2(1-z)}
\label{dipolsplit}
\end{equation}
and then integrate the first term from 0 to 1. (This division corresponds to the separation used in the dipole cascade model \cite{dipol}, but the same result is obtained for any other symmetric separation.)
The result in eq.~(\ref{lgrans}) implies that the effective phase space for the $l_i$ integrations in e. g. eq.~(\ref{dglap}) is reduced. It corresponds to an effective limit $\ln 1/z_i = l_i - l_{i-1} > a$ instead of $l_i - l_{i-1} > 0$, with $a \approx 11/12$. This means that instead of the factor $l^N/N!$, the $l_i$ integrals give the result  $(l-aN)^N/N!$. Thus in eqs. (\ref{fixsumma}) and (\ref{Gslut}) we should make the substitutions
\begin{equation}
l^N/N! \rightarrow (l- a N)^N/N!.
\label{nlo}
\end{equation}

Such a substitution was discussed in ref. \cite{LDC} and shown to imply a significant reduction in the value of $\lambda$. Using Sterling's formula it is easy to show the following relation between the old value $\lambda$ and the new, called $\lambda'$:
\begin{equation}
\ln \frac{\lambda}{\lambda'} = a \lambda'.
\label{lambda'}
\end{equation}
Thus for $a=11/12$ and $\lambda=0.5$ we obtain $\lambda' \approx 0.35$.

In principle the effect is, however, not restricted to a reduction of the exponent $\lambda$. The factor  $(l-a N)^N/N!$ also means that there is a strict upper limit, $N < l/a$, to the sum over $N$ in eqs.~(\ref{fixsumma}) and (\ref{Gslut}). The increase for very large $k_\perp$-values is due to contributions from large values of $N$, and therefore this restriction implies that the distribution in $\kappa$ will turn over for large values of $\kappa$. In the large $k_\perp$-limits of eqs.~(\ref{dglapasfix}) and (\ref{Gdglap}) the dominating contributions come from $N$-values around $N_{max}=\sqrt{\bar{\alpha} l \kappa}$ (fixed coupling) or $N_{max}=\sqrt{\alpha_0 l u}$ (running coupling). Therefore there is a suppression for $k_\perp$-values such that $N_{max} > l/a \approx l$, which means $\kappa>l/\bar{\alpha}$ for fixed coupling and $u =\ln (\kappa/\kappa_c) > l/\alpha_0$ for running coupling. For running coupling and small $x$ this corresponds, however, to unrealistically large $k_\perp$-values. Consequently the dominant effect of these non-leading terms is just a reduction of the BFKL exponent $\lambda$, in accordance with eq.~(\ref{lambda'}). (This reduction also implies that the boundary between the the BFKL and DGLAP regimes becomes somewhat modified. Thus the ratios in eqs.~(\ref{fixgrans}) and (\ref{grans}) are reduced by a factor $\frac{\lambda'}{\lambda(1+a\lambda')}$.)

\section{Laplace transforms}

Taking the Laplace transform of the sum in eq.~(\ref{fixsumma}), relevant for a fixed coupling, does not give a simple expression. However, for a running coupling it is easy to calculate the Laplace transforms of the expression in eq.~(\ref{Gslut}). For the single Laplace transform we find
\begin{equation}
\widetilde{G}(\omega,u) = \int d l e^{-\omega l} G(l,u) = \frac{1}{\omega - \lambda} \cdot e^{\frac{\alpha_0 u}{\omega}}.
\label{singlelaplace}
\end{equation}
Thus we have a simple pole for $\omega = \lambda$ and an essential singularity for $\omega = 0$. The simple pole is in contrast to the case with a fixed coupling, where the factor $(\ln 1/x)^{-1/2}$ in eq.~(\ref{bfklasfix}) implies a cut for $\omega < \lambda$.   Denoting the $m$th term in the series in eq.~(\ref{Imsumma}) by $G_m(l,u)$ we also find
\begin{equation}
\widetilde{G}_m(\omega,u) = \frac{1}{\omega} (\frac{\lambda}{\omega})^m \cdot e^{\frac{\alpha_0 u}{\omega}}.
\label{laplacem}
\end{equation}
Here we see how the contributions for different $m$-values add up to the pole at $\omega = \lambda$, and thus to the power-like increase $\sim x^{-\lambda}$ for small $x$.

Including the non-leading effects from the substitution in eq. (\ref{nlo}), the result is somewhat more complicated, but the singularity at $\omega = \lambda$ is still a simple pole and not a cut.

Finally we note that the double Laplace transform of eq~(\ref{Gslut}) has a very simple form (note that this is the Laplace transform in $u=\ln \ln k_\perp^2$ and not in $\ln k_\perp^2$)

\begin{equation}
\widetilde{G}(\omega,\gamma) = \int d l \, e^{-\omega l} \int d u \, e^{-\gamma u} G(l,u) = \frac{1}{(\omega - \lambda) (\gamma - \frac{\alpha_0}{\omega})}. 
\label{doublelaplace}
\end{equation}

\section{Summary}

A simple model is presented, which interpolates smoothly between the DGLAP and BFKL regimes. It is based on the CCFM and LDC models and incorporates the most essential features of these models. Although not meant for quantitative analyses, it provides an intuitive picture of the transition between the two domains.

To leading order in $\ln 1/x$ we get for a constant coupling 
\begin{equation}
 {\cal F}(x,k_\perp^2) \approx \bar{\alpha} \sum_N \frac{(\bar{\alpha} l)^N}{N!} \frac{(\kappa+N\delta)^N}{N!}
\end{equation}
where ${\cal F}$ is the non-integrated gluon structure function, $l=\ln 1/x$, $\kappa=\ln k_\perp^2$, and $\bar{\alpha} = 3\alpha_s/\pi$. The parameter $\delta$ should be of order 1, and the BFKL power is given by $\lambda = e \delta \bar{\alpha}$, which is very close to the leading order BFKL result. The transition between the DGLAP and BFKL regimes occurs for the fixed ratio
\begin{equation}
\frac{\kappa}{l} =  \frac{\ln k_\perp^2}{\ln 1/x} \approx e \,\bar{\alpha}.
\end{equation}

For a running coupling we get instead
\begin{equation}
 \frac{\kappa}{\alpha_0} {\cal F} \equiv G\approx  
\sum_N \frac{(\alpha_0 l \Delta)^N}{N!} \sum_{k=0}^N \frac{(u/\Delta)^k}{k!} 
\end{equation}
where $u=\ln \ln k_\perp^2 - \ln \ln k_{\perp cut}^2$ and $\alpha_0$ is defined by the relation $3\alpha_s/\pi = \alpha_0/\kappa$. The parameter $\Delta$ is sensitive to the cutoff for small $k_\perp$ (see Fig.~\ref{fig:delta} ), and related to the BFKL power by the equation $\lambda=\alpha_0 \Delta$. In this case the transition occurs for 
\begin{equation}
\frac{u}{l} =  \frac{\ln \ln k_\perp^2}{\ln 1/x} \approx \frac{\lambda^2}{\alpha_0}.
\end{equation}

Some non-leading effects in $\ln 1/x$ can be taken into account by replacing the factor $l^N$ in eqs. (\ref{fixsumma}) and (\ref{Gslut}) by $(l-a N)^N$ with $a\approx11/12$. The result is mainly just a smaller value for the BFKL exponent $\lambda$ (in agreement with results in refs.~\cite{LDC,BFKL}).

While the Laplace transform for a fixed coupling has a cut for $\omega < \lambda$, for a running coupling the singularity at $\omega = \lambda$ is only a simple pole.

\section{Acknowledgments}
We would like to thank Bo Andersson and Leif L\"onnblad for many valuable discussions.


\end{document}